\documentstyle[aps]{revtex}
\begin{document}

\title{
Topological phase transition in 2D quantum Josephson array
}

\author{A.I. Belousov and Yu.E. Lozovik\cite{e-mail}}

\address{Institute of Spectroscopy, Russian  Academy  of  Sciences,\\
142092, Troitsk, Moscow region, Russia. }

\maketitle

\begin{abstract}
Path Integral Quantum Monte Carlo simulation is used to study
thermodynamic properties and a phase diagram of 2D quantum
Josephson array,
described by 2+1 XY model.  The helicity and vorticity moduli,
correlation  function  of  phases and other
characteristics of the system as functions of quantum parameter $q$
and temperature $T$ are studied ($q=2e/\sqrt{JC_0}$, $J$ is the
Josephson coupling  constant,  $C_0$ is  the  intragrain capacitance).
Quantum fluctuation induced superconductor -- normal phase transition
is studied in detail through the use of behavior of above-mentioned
quantities.  No discontinuous or reentrant phase transition  in  $q -
T$  plane  is found.  Analysis of the vorticity and the renormalized
coupling constant leads to the conclusion that the whole line of
phase transition is        of        Kosterlitz         -
Thouless type.  \end{abstract}

\section{Introduction}
\label{introduction}
Ordering phenomena in 2D Josephson arrays has elicited considerable
theoretical and experimental interest (see \cite{Abeles} - \cite{Choi}
and references therein).  The phase transition in {\it classical} system
when capacitances are neglected is generally believed to be of
Kosterlitz - Thouless type. Taking into account intra- and intergrain
capacitances leads to a variety of interesting phenomena, which are
governed by the competition between the Josephson tunneling and the
charging energy of the grain. This charging effect leads to the breaking
down (at appropriate sizes of granules) of superconductivity even at
zero temperature.

A major outcome of the simulation of periodic array of ultrasmall granules
\cite{JosePhRev} was the observation of a transition to a {\it new coherent
state}. At sufficiently small capacitances it was shown to be two
"ordering" phenomena: the system undergoes at higher temperatures
Kosterlitz - Thouless transition from the disordered phase into the
superconducting one, and upon further decreasing of temperature a
quantum induced first order transition to another superconducting phase
with finite but diminished superfluid density.

It is important to analyze a possible modification of mechanism and
type of the phase transition when quantum effects begin to play an important
role.  Because the critical behavior of the quantum system is similar to
that of the thick classical anizotropic film
(and is described in terms of 2+1 XY model),
the Kosterlitz--Thouless phase transition scenario may be in principle
replaced (for the finite size system) by behavior, inherent to 3D anizotropic
superconductor \cite{Doniach},\cite{Akopov,Kultanov}.
By introducing a Hikami-Tsuneto (HT) length scale defined by the ratio
of the inter($J_{\bot}$) and intra($J_{\|}$) - plane coupling constants
\cite{Hikami}
$P_{HT}=\sqrt{ J_{\bot} / J_{\|} } = P / (q \beta J)$
($P$ - is the thickness of the anizotropic classical film,
$\beta=1/k_{b}T$) it can be easily seen, that at distances
$|\vec{r}_{i} - \vec{r}_{j}| \le P / P_{HT}$
the behavior of the system has 3D character with a
corresponding correlation function \cite{Akopov} and vortex characteristics,
while at $|\vec{r}_{i} - \vec{r}_{j}| > P / P_{HT}$
the system has 2D behavior.
Thus, it is interesting to analyze
topological excitations responsible for phase transition in the region of
strong quantum fluctuations.

The purpose of this Letter is {\it ab initio} studying properties and the
phase diagram of the 2D quantum Josephson array.
By path integral Monte Carlo simulations (PIMC) we will calculate
different physical properties (helicity and vorticity moduli, correlation
function, and vortex moment)
on the plane $\{q,T\}$, where the quantum parameter $q = 2e /
\sqrt{J C_0}$ is responsible for the strength of zero - point fluctuation of
phases, $T = k_b T / J$ is the dimensionless temperature of the system.
From the analysis of above-mentioned quantities
we will show, that at distances
larger than $P / P_{HT}$ the system experiencing 2D
Kosterlitz--Thouless - like
phase transition with vortex unbinding and thus can be regarded as 2D
classical system with a coupling constant renormalized by quantum
fluctuations.

\section{Model and calculating properties.}
\label{model}

At sufficiently low temperatures fluctuations of moduli of the
superconducting  order parameter can be neglected and the
Hamiltonian of the 2D system of $N^2$ granules can be written as
\begin{eqnarray*}
\hat H = \frac{2e^2}{C_0} \sum\limits_{i} (\hat n_i - n_0)^2  +
J \sum\limits_{<i,j>}(1-\cos{(\hat\varphi_i - \hat\varphi_j)})
\end{eqnarray*}
The first term of the Hamiltonian is related to the Coulomb charging energy
($C_0$ is the intragrain capacitance;
We suppose that intergrain capacitances  $C_1 \gg C_0$
and their effect may be neglected). $\hat n_i - n_0$ stands for the
deviation of the number $n_i$ of effective bosons (Cooper pairs)
from its equilibrium value $n_0 = <\hat n_i>$.
Josephson coupling energy $J$ in the second term is assumed the same
for all nearest - neighbor pairs $<i,j>$ of islands. Periodical boundary
conditions are used.
Phases $\varphi_i$ of the order parameter are chosen to be cyclic
variables: $\varphi_i \in [0,2\pi),\;\; i = \overline{1 \ldots N^2}$.

If the relative fluctuations of the number of Cooper pairs
are small then the particle number operator $\hat n_i - n_0$ can be chosen
as one conjugate to the 'phase' operator \cite{Bruder,BelousJPC}:
$\hat n_i - n_0 = \jmath \partial / \partial \phi$. This lead to the
Hamiltonian of 2D quantum XY model:
\begin{eqnarray}
\hat{H}=-\frac{2e^2}{C_0}
\sum_{i=1}^{N^2}
\frac{\partial^2}{\partial \varphi_i^2}+
J\sum_{<i,j>}
(1-\cos(\varphi_i-\varphi_j))
\label{_hamiltonian}
\end{eqnarray}

After the simple discretisation procedure for path integral,
in the so - called primitive  form of PIMC (see e.g. \cite{Ceperley}),
one can represent properties of an initial 2D quantum system
by fictitious 3D
classical system formed by the $P$ -- times multiplication of the initial one.
It is described
by Boltzmann partition function $w$ with an effective potential $V_{eff}$:
\begin{eqnarray}
w(\varphi_1^0 \ldots \varphi_{N^2}^{P-1};\beta) = \exp{(-\beta V_{eff})},
\nonumber
\\
\beta V_{eff}(\varphi_1^0 \ldots \varphi_{N^2}^{P-1}) =
\sum_{p=0}^{P-1}{\left[ \sum_{i=1}^{N^2}
\frac{T P \left(\varphi_i^{p+1}-\varphi_i^{p}-2\pi n_i^* \right)^2}
{2 q^2}
+\frac{1}{T P}\sum_{<i,j>}
(1-\cos(\varphi_i^p-\varphi_j^p))\right]}
\nonumber
\\
\varphi_i^P = \varphi_i^0.
\label{V_eff}
\end{eqnarray}
an integer $n_i^*$ provides the minimum of
$|\varphi_i^{p+1}-\varphi_i^{p}-2\pi n_i^*|$ at some fixed
$\varphi_i^{p+1}$, $\varphi_i^p$ and approximately takes into account
that phases $\varphi_i^p$ are cyclic variables.
An equilibrium value of some observable
$\hat{O}$
then can be found as:
\begin{eqnarray}
\left<{\hat{O}}\right> \approx
\left<{\hat{O}}\right>_P =
\frac{
\int\limits_0^{2\pi} \ldots \int\limits_0^{2\pi}
O(\varphi_1^0,\varphi _2^0, \ldots \varphi _{N^2}^0)
w(\varphi_1^0 \ldots \varphi_{N^2}^{P-1};\beta)
d\varphi_1^0 \ldots d\varphi_{N^2}^{P-1}
}
{
\int\limits_0^{2\pi} \ldots \int\limits_0^{2\pi}
w(\varphi_1^0 \ldots \varphi_{N^2}^{P-1};\beta)
d\varphi_1^0 \ldots d\varphi_{N^2}^{P-1}
}
\label{A_P}
\end{eqnarray}
In principle we can substitute the fictitious classical system in hand into
the anizotropic 3D XY one with the coupling constants
$J_{\|} = J/P,\;\;J_{\bot}=P/(q^2 J \beta^2)$. The anizotropy of this
system is defined by the ratio $J_{\bot} / J_{\|} = 2/\tau_1$, were the
parameter $\tau_1 \ll 1$
is responsible for the accuracy of the
primitive discretisation procedure (see below).
So, one can imagine, that the behavior of the
anizotropic 3D XY system
$N \times N \times P$ with a {\it constant anizotropy}
is studied at different
points $\{q,T\}$ of the phase diagram and, consequently, with different
thickness $P$.

Properties of the system were explored in approximately 150 points
belonging to the region $ 0.1 \le q \le 2.6,\,\,\,\,\, 0.02 \le T \le 1.4 $.
We used the multigrid algorithm (unigrid modification \cite{Sayer} was used).
More precisely, we used the $W_{10}$ cycle
(i.e. the W cycle with 1 presweep and no postsweeps)
in the "classical region" $q < 0.5$ while the $V_{10}$ cycle was found
to be more efficient at $q > 0.7$.
The Trotter number $P$ at each point of phase plane was fitted in order to
the relative discretisation error of Feynman integral being smaller that
$3 \%$.
This type of error is determined by the discretisation parameter
$\tau_1 = 2 q^2 / T^2 P^2$, its value was $0.001 < \tau_1 < 0.07$
at all considered points.
Note, that the errors related with partial neglecting of the periodical boundary
conditions for temperature density matrix
(the appropriate parameter is
$\tau_2 = 2\exp{ \left( -2 \pi^2 T P/q^2 \right) }$)
are small with such Trotter numbers and are of order of $2\%$
of the calculated mean values.

The following quantities where measured at all above-mentioned points
of the $\{q,T\}$ plane:

{\bf 1) The modified Lindeman ratio.}

This quantity is the generalization of the well-known Lindeman's
rms displacement of particles (phases).
The Lindeman criterion states that solid melts when the
rms displacements become larger than some universal part $\delta_c$
of the interparticle distance
($\delta_c$ is generally assumed to be $\approx 0.15$).
For 2D system this value is known
to diverge logarithmically with the system size. This
does not enable us to take advantage of this universality
in determining the temperature of phase transition.
Instead of rms displacement one can use the modified Lindeman ratio
\cite{Bedanov}:
\begin{eqnarray}
\delta _l =
\left< \frac{1}{2 N^2} \sum_{<i,j>\,\,,l}
\left\lfloor \varphi _i - \varphi _j \right\rfloor_{[-\pi,\pi)}^2
\right>^{1/2}
\label{Lozovik}
\end{eqnarray}
The sum in the formula (\ref{Lozovik}) is taken
over such pairs of granules $<i,j>$ that
the vector drawn from the granule $i$ to $j$ is equal to
$(0,l)$ or $(l, 0)$. In view of
periodic boundary conditions the number of such pairs is equal to $2N^2$.
Here and henceforth $\left\lfloor f \right\rfloor_{[a,b)}$
denotes the reduction of $f$ to the interval $[a,b)$.

{\bf 2) The helicity modulus.}

This quantity gives the information about the "rigidity" of the system
perturbed by imposed gradient
$\delta \phi(m,n) = \vec{k} \vec{r}(m,n) = k_x m + k_y n \;$,
$\vec{r}(m,n) \in [1,N] \times [1,N]$.

Helicity modulus \cite{Ohta} (see also  \cite{Minhagen}) can be
defined as
$
\gamma_{ab} =
\frac{1}{J N^2}
\left.\frac{\partial ^2 F}{\partial k_a \partial k_b}\right| _{k \to 0}
$
where $F = -k_b T ln(Z)$ is free energy of the system.
It is worth-while to note, that the classical expression
(see e.g. \cite{Papricash}) for $\gamma$  is not correct for quantum
system.
The reason is the same as that pointed out in the review
\cite{Ceperley} when calculating superfluid density: operators
$d \hat H/d k$ and $e^{-\beta \hat H}$ do not commute.
After some algebra,
the following expression for the helicity modulus of the quantum XY model
can be obtained \cite{JosePhRev}:
\begin{eqnarray}
\gamma_{ab} = \gamma \delta_{ab},
\nonumber
\\
\gamma = \frac{1}{N^2}
\left<
\sum\limits_{m,n=1}^{N} \cos{(\varphi^0 (m+1,n) - \varphi^0 (m,n))}
\right>_P -
\nonumber
\\
\frac{1}{N^2 P T}
\left< \sum\limits_{p=0}^{P-1}
\sum\limits_{m,n,k,l=1}^{N}
\sin{(\varphi^0(m+1,n) - \varphi^0(m,n))}
\sin{(\varphi^p(k+1,l) - \varphi^p(k,l))}
\right>_P
\label{helicity}
\end{eqnarray}

On assuming that $\varphi^p(m,n) = \varphi^0(m,n)$,
one can obtain the well-known classical expression \cite{Papricash}:
\begin{eqnarray}
\gamma_{cls} = \frac{1}{N^2}
\left<
\sum\limits_{m,n=1}^{N} \cos{(\varphi^0(m+1,n) - \varphi^0(m,n))}
\right>_P -
\frac{1}{N^2 T}
\left< \left[ \sum\limits_{m,n=1}^{N}
\sin{(\varphi^0(m+1,n) - \varphi^0(m,n))}
\right]^2
\right>_P
\label{helicity_cls}
\end{eqnarray}
Being calculated, the "classical helicity modulus" can be regarded as
a low bound to the true one (\ref{helicity}).
It occurs that the classical expression
gives rather a good approximation of (\ref{helicity}) almost for all
the superconducting region of the $\{q,T\}$ plane excluding $q > 2$ and small
temperatures $T < 0.05$ where the number  of effectively
uncorrelated planes $P / P_{HT}$ is large enough and the imaginary -- time
correlator in (\ref{helicity}) is far from constant.

{\bf 3) The space correlation function.}

The space correlation function $g(l)$ of phases can be introduced as:
\begin{eqnarray}
g(l) = \left<
{ \frac{1}{n_l}
\sum\limits_{|\vec{r}_i - \vec{r}_c|=l}
  \cos{(\varphi_{i}-\varphi_c)}
}\right>
\label{Corr_function}
\end{eqnarray}
where $n_l$ is the number of granules at the
distance of $l$ ($0 \le l \le \left\lfloor{\frac{N}{2}}\right\rfloor - 1$)
from some fixed one $\varphi _c$.

{\bf 4) The density of open vortices.}

If there are $2n$ open vortex threads in the system
$[0,N] \times [0,N] \times [0,P]$ (see e.g. \cite{Shenoy})
then the density of vortices $\rho_v$ is defined as:
\begin{eqnarray}
\rho_v = \frac{n}{N^2}
\label{DensVor}
\end{eqnarray}

{\bf 5) "Vorticity modulus".}

This quantity is the measure of the response of the system to
an isolated vortex (open vortex thread in our case)
introduced into the system and may be
defined as \cite{Kicuchi}:
\begin{eqnarray}
v_{DAP-P} = \frac{ F_{DAP} - F_{P} }{ \ln{N} }
\label{Vorticity}
\end{eqnarray}
Here $F_{P}$ is free  energy of the system under periodic ('P')
boundary conditions when the net topological charge is equal to zero
and there always appear only equal numbers of vortices and antivortices.
$F_{DAP}$ is free energy of the system under
diagonally -- antiperiodic ('DAP') boundary conditions (see below)
when there is at least one excess vortex in the system.

Variational upper and lower bounds on free energy differences (\ref{Vorticity})
can be obtained {\it via} Bogoliubov - Gibbs variational method
(see e.g. \cite{Hogerson}).
To find upper bound, for example, the coupling constant at the boundary
$J_t^{b}$, and consequently the Hamiltonian, is changed slowly over the course
of the simulation (where the parameter $t=\overline{0 \ldots 2M}$ marks
the Hamiltonian updates)
from its initial value $J_0^b = J$ to zero $J_{M}^b = 0$ (free boundary
conditions), system being at periodic boundary conditions.
At this the very moment $t = M$ the type of boundary
conditions is switched to DAP i.e.
$$
\varphi^p(m,0) = \left\lfloor \pi + \varphi^p(N-m+1,N) \right\rfloor_{[0,2\pi)},
\;\;
\varphi^p(m,N+1) = \left\lfloor \pi + \varphi^p(N-m+1,1) \right\rfloor_{[0,2\pi)},
$$
$$
\varphi^p(0,n) = \left\lfloor \pi + \varphi^p(N,N-n+1) \right\rfloor_{[0,2\pi)},
\;\;
\varphi^p(N+1,n) = \left\lfloor \pi + \varphi^p(1,N-n+1) \right\rfloor_{[0,2\pi)}.
$$
Subsequent increasing of the coupling constant on the boundary up to its
initial value $J_{2M}^b = J$ led system to the final state $t = 2M$ with
at least one free vortex presented.

At each step $t$ the system was heated for
$\sim 400 N^2$ Monte Carlo (MC) steps (one MC step is the $V$ cycle or
the $W$ cycle).
Measuring of the work $\delta A_t$ of altering the coupling constant
from $J(t)^b$ to $J(t+1)^b$ required some more $\sim 400 N^2$ MC steps.
The energy difference caused by alteration of the coupling constant at the
boundary $J(t)^b \to J(t+1)^b$ contributes to the work as:
$$
\delta A_t = \left< H(J_{t+1}^b) - H(J_{t}^b) \right>_{J_t^b} =
\left<
J_{t+1}^b \sum\limits_{b}
(1-\cos(\varphi_i-\varphi_j)) -
J_{t}^b \sum\limits_{b}
(1-\cos(\varphi_i-\varphi_j))
\right>_{J_t^b}.
$$

The upper estimation of free energy difference
(\ref{Vorticity}) is calculated as the total work required for
changing of boundary conditions from P to DAP:
\begin{eqnarray*}
F_{DAP} - F_{P} \le \sum\limits_{t=0}^{2M} \delta A_t^{P \to DAP}
\end{eqnarray*}

The calculation of the lower bound of free energy differences required
the analogous transformation from the DAP to P. Similarly, we obtain:
\begin{eqnarray*}
F_{DAP} - F_{P} \ge -\sum\limits_{t=0}^{2M} \delta A_t^{DAP \to P}
\end{eqnarray*}

In present work, we found that a good choice of slowly varying
coupling constant at the boundary is
$$
J_t^b = J \cos^2{(\pi t / 2M)}.
$$
The concrete number of Monte Carlo steps for heat and
measure as well as the total number $2M$ of Hamiltonian updates
where fitted to provide the accuracy (it can be estimated as the difference
between the upper and the lower bounds of the vorticity modulus)
needed for the estimation (\ref{Vorticity}).

There is another way of looking at the vorticity modulus.
Let us assume that the system is placed to a magnetic field {$\bf H$}
such that:
$$
(A_x,A_y,A_z) = \frac{\Phi_0 s}{2 \pi (x^2+y^2)} (-y,x,0), \;\;\;\;\;
\Phi_0 = \frac{2\pi c \hbar}{e}
$$
$$
(H_x,H_y,H_z) = \frac{s \Phi_0  \delta(x^2+y^2)}{\pi} (0,0,1)
$$
Let the origin of the frame of reference
be in the center of the array. Consider
the free energy of the system as a function of the strength $s$ of the field.
In the limit of small $s$ one can write
\begin{eqnarray*}
F(s) = F(0) + \frac{J V}{2} s^2 + O(s^3)
\end{eqnarray*}
The linear response
$J V = \left. \partial^2 F(s) / \partial s^2 \right|_{s=0}$
of the system to an isolated vortex at the origin takes the form:
\begin{eqnarray}
V =
\left<
\sum\limits_{m,n=1}^{N} \left\{
\Lambda_x^2(m,n) \cos{(\varphi^0(m+1,n) - \varphi^0(m,n))} + \right. \right.
\left. \left.
\Lambda_y^2(m,n) \cos{(\varphi^0(m,n+1) - \varphi^0(m,n))}
\right\}
\right>_P -
\nonumber
\\
\frac{1}{TP^2}
\left<
\left[
\sum\limits_{p=0}^{P-1}
\sum\limits_{m,n=1}^{N} \left\{
\Lambda_x(m,n) \sin{(\varphi^p(m+1,n) - \varphi^p(m,n))} + \right. \right. \right.
\left. \left. \left.
\Lambda_y(m,n) \sin{(\varphi^p(m,n+1) - \varphi^p(m,n))}
\right\}
\right]^2
\right>_P,
\label{V}
\end{eqnarray}
\begin{eqnarray*}
\Lambda_x(m,n) = \arctan{\left(x_{m,n}/y_{m,n}\right)}  -
\arctan{\left(x_{m+1,n}/y_{m,n}\right)},
\\
\Lambda_y(m,n) = \arctan{\left(y_{m,n+1}/x_{m,n}\right)}  -
\arctan{\left(y_{m,n}/x_{m,n}\right)},
\end{eqnarray*}
where $(x_{m,n},y_{m,n})$ stands for the position of the granule
$(m,n)$ in the particular frame of reference.

Analogous expression was used in studying the phase diagram
of the Heisenberg antiferromagnet \cite{Southern}.
As have been pointed out, the $V$ contains both a core energy $E_c$
and a part which is proportional to $\ln{N}$:
$$
V(N) = E_c/J + v_H \ln{N}
$$
The size dependent part can be extracted then by using the results
obtained for systems of size $N_1$ and $N_2$ \cite{Southern}:
\begin{equation}
v_H = \frac{V(N_2) - V(N_1)}{\ln{(N_2/N_1)}}
\label{v}
\end{equation}

\section{Results and discussion.}
\label{results}
Figures~2a,b show behaviour of the modified Lindeman ratio (\ref{Lozovik})
in the explored region of parameters $\{q,T\}$. From the results of
calculations one can see that neither in the 'classical' region ($q < 1$) nor
in the quantum one, this quantity does not have any points of abrupt changes
that could have revealed the line $T_c(q)$ of phase transitions.
But it is hoped that the values $\delta_l^*$ at the point $\{q_c,T_c\}$
of phase transition may be universal for all 2D system of the same symmetry.
For example, in the case of classical system ($q=0$) we have
$\delta_1^* = \delta_1(0,0.89) \approx 0.81, \;\;
\delta_3^* = \delta_3(0,0.89) \approx 1.06$. It should be
particularly emphasized that the abovementioned quantities $\delta_l$
have another values $\delta_l^*$ at the point $\{q,T\}=\{q_c,0\}$
of quantum phase transition
(this universal values, if any, are the same for analogous quantum system).

Disappearance of the helicity modulus testifies to the disordering of phases
in the system.
Let us analyze the behavior of $\gamma(q)$  (see Fig.~3a).
As the quantum parameter $q$ increases, the value of $\gamma$ decreasing
slowly up to some critical $q^*$ rapidly falls to zero in a narrow
interval $(q^*, q^* + \Delta q)$.
Analogous behavior of this quantity takes
place along all lines $q = const.$ Really, at $q = 0$ there is a sharp drop of
the quantity $\gamma(T)$ at $T^* \approx 0.9$, associated with the Kosterlitz --
Thouless phase transition in the classical system.
At Figure~3b examples of such classic -- like behavior of $\gamma(T)$ for
different values of the quantum parameter $q$ are presented.
From the Figures~3 one can see, that the curves $\gamma(q)$ and $\gamma(T)$
are very gently sloping in the region $q > 1.0$.

In the Fig.~3a (solid diamonds)
the results of calculations of the helicity modulus at
$q = 0.5$ and at low temperatures $T<0.1$ are given.
In this the very region of temperatures, at $T \le 0.03$
authors of the work \cite{JosePhRev} marked a discontinuity in helicity
modulus and specific heat as functions of temperature.  This phenomena was
associated with the first order phase transition due to quantum fluctuations
of phases.
Presented results correspond to the system
$N \times N = 10 \times 10$ (as well as in the \cite{JosePhRev}),
but our discretisation errors are much smaller:
Trotter number $P$ was $P|_{T = 0.02} = 128$
($\tau_1 \approx 0.07$ compared to $\tau_1 \approx 0.25$ in the
\cite{JosePhRev}). We see that within the limits of statistical
errors any disordering phenomena are absent in this region $\{q,T\}$.
No reentrance (or discontinuity) are observed
(see Fig.~3b) also at larger values of the quantum parameter $q$,
i.e even when a strength of quantum fluctuations is rather great.
It is worth-while to note, that the {\it classical} expression
for the helicity modulus (\ref{helicity_cls})
is incorrect at this region of dimensionless parameters of the system.
The deviation of the value of $\gamma$ calculated by the adequate expression
(\ref{helicity}) from that calculated by classical expression
may have the order of $\gamma$
both at small temperatures $T < 0.05$ (at $q \sim 0.5$) and at
large values $q > 2$ of the quantum parameter,
to lead to the appreciable altering of the phase diagram.
This suggestion is supported by results presented at Figures~3a,b.

As pointed out in \cite{JosePhRev}, the helicity modulus
(the superfluid density) tends to a finite value of order one as
$T \to 0$ (we have $\left.  \gamma \right|_{T=0} \sim 0.87$ at
$q \sim 0.5$).  So, we can apply the spin wave consideration to
the anizotropic system $N \times N \times P$ in this region of
temperature.  Through the appropriate scale transformation we
obtain the izotropic system $N \times N \times P/P_{HT}$ with
coupling constants $J_{\|} = J_{\bot} = P_{HT} / P$,
$P_{HT} = PT/q$. Note, that the thickness of
this system ($T=0.02,\;q=0.5,\;N=10$) is three times larger than
its width: $P/P_{HT} =q/T \sim 25 > N$ and hence we are to observe
3D - 1D crossover behavior at this parameters of the system.

The quantum $N \times N$ Josephson array is known to have 3D - like
behavior below the line $T = q / N$ \cite{Doniach},\cite{Akopov,Kultanov}.
In the above region we are to observe 2D behavior and, hence,
the boundary of the global superconductive state in $q - T$ phase
diagram ($T > q / N$) is expected to be a line of Kosterlitz -- Thouless
transitions with temperatures $T_c = T_c(q)$ defined {\it via.}
the renormalized by quantum fluctuations coupling constant $J(q)$.

Insufficient accuracy of calculations and small system sizes does not
permit, as a rule, to check for agreement
between theoretical Kosterlitz -- Thouless critical exponents
and calculated ones.
This implies that the conclusion about a type and a character of the phase
transition may be done on the basis of some other specific properties.
The vorticity modulus is
one of them for topological Kosterlitz -- Thouless phase transition and
is proportional to excess free energy due to an excess vortex.
In terms of the 2+1 XY model we may consider that this quantity is the
measure of the ability of the $N \times N \times P$ system to form an
isolated unclosed vortex thread. From the definition of this quantity
(\ref{Vorticity}) it follows that the vorticity modulus $v_{DAP-P}$ being
positive in the ordered phase, should vanish at the disordered one.
Dependencies $\left. v_{DAP-P}(T) \right|_{q=const.}$
for the system $N \times N = 12 \times 12$
and different values of the quantum parameter are presented in Fig.~4a
(the dependencies $v_H(T)$ and $v_H(q)$, see (\ref{v}), are given
at Figures~5a,b).
Shapes of curves $\left. v_{DAP-P}(T) \right|_{q=const.}$
for classical ($q = 0$) and
quantum regions are very much alike.  The tendency of decreasing the phase
transition temperature with increasing $q$ correlates with the behavior of
the vorticity modulus:  the temperature $T_v$ at which the vorticity changes
sign, i.e. it is advantageous to born an excess vortex thread,
is shifted to the lower temperatures as quantum parameter $q$ is increased.
This observation is in
agreement with the results of calculations of the helicity modulus {\it vs.}
temperature at different $q$ (see Fig.~3b).  The dependence of the vorticity
{\it vs.}
$q$ at $T = 0.5$ is given at Fig.~4b.  As was noted by authors of
\cite{Kicuchi}, the temperature $T_v$ decreases only weakly with the system
size.  This is the reason, by which we may use data on calculating vorticity
modulus for the system $12 \times 12$ together with other results for the
system $20 \times 20$.

The point $\{q_v,T_v\}$ at which the vorticity modulus (\ref{Vorticity})
changes sign
may be considered as one at which the mean separation of free vortices
matches the system size \cite{Kicuchi}. So it would be interesting to compare
points $\{q_v,T_v\}$ with those of great variance of the correlation
length of phases.
Let us suppose that an empirical formula for the correlation function
(\ref{Corr_function})
$g(r) \sim \exp{(-r / \xi)}$ takes place.
Experimental data on $\xi$ as a function of $q$ (or $T$)
is presented at fig.~4b (Fig.~4a).
It may be seen, that points $\{q_v,T_v\}$
are belongs to the regions of considerable increase of the empirical
parameter $\xi$ which may be associated then with the correlation length
of the system.

An analysis of the helicity and the voricity moduli
make it possible to consider the point $\{q,T\} = \{2.2,0.5\} = \{q_c,T_c\}$
as the point of phase transition.
Let us explore the region $q \sim 2,\; T \sim 2$
of strong quantum fluctuations near this point.
The corresponding 3D effective izotropic system in this region has dimensions
$N \times N \times P/P_{HT} \sim 20 \times 20 \times 4$ and can be
regarded as 2D one at distances $|\vec{r}_i - \vec{r}_j| > 4$.
Particles of this 2D system,
interacting {\it via} the coupling constant $J(q,T)$ renormalized
by quantum fluctuations, can be
associated with open threads (of appropriate topological charge).
To determine the renormalized coupling constant
$J = J(q,T)$ of the 2D system of charges at some point $\{q,T\}$ one
can analyze the density $\rho_v$ (\ref{DensVor})
of open threads in the system.
When $\rho _v << 1$ one can write \cite{Thouless}:
\begin{eqnarray*}
\rho _v \sim \frac{\pi e^{-2 \pi^2 \beta J(q,T)}}
{\pi \beta J(q,T) -1} \left( 1 - N^{2 - 2\pi \beta J(q,T)}\right).
\end{eqnarray*}
After $\rho _v(q_c,T_c)$ was being calculated, we compared the experimental
renormalized coupling constant $J(q_c,T_c)$
with that, obtained by Kosterlitz -- Thouless criteria:
$J_c = T_c / 0.93 \approx 0.54$. Rather a good conformity
between both the results testify to the correctness of
the Kosterlitz -- Thouless picture of phase transition in this
quantum region (more detailed analysis of quantities described the
vortex structure of the system will be given elsewhere).

Another argument is the universal jump of the helicity modulus
in this the very point $\{q_c,T_c\}$. The point $q_{j}$ of the universal
jump is that of intersecting of the line
$\left. 2T / \pi \right|_{T=0.5} = 1/\pi$
with the plot of the helicity modulus $\left. \gamma(q) \right|_{T=0.5}$
(see Fig.~3a) and is equal to $q_{j} = 2.15 \pm 0.05$.

The resulting phase diagram (see Fig.~1) is in sufficiently good agreement
with that calculated from Kosterlitz -- Thouless scenario through the use of
self-consistent harmonic approximation (SCHA) \cite{Akopov,Verzak}.

To conclude, we have analyzed properties of 2D Josephson array
at different quantum parameters $q$ and temperatures $T$.
The phase transition in all the region (above the region of crossover)
have been found to be of Kosterlitz -- Thouless type.
No reentrance or discontinuity phenomena have been found.

\vspace{0.5cm}

{\bf Acknowledgments}.
We would like to thank N.K. Kultanov and S.A. Verzakov for
valuable discussions. The work was supported by Russian Foundation for Basic
Research and by program "Physics of Solid Nanostructures".

\vspace{0.5cm}

Fig.~1. \\
The phase diagram of 2D quantum Josephson array.
(1) Results of present calculations.
(2) Results of Ref. \cite{JosePhRev}.
(3) SCHA renormalized Kosterlitz -- Thouless temperatures \cite{Verzak},
\vspace{0.3 cm}

Fig.~2a. \\
Modified Lindeman ratio $\delta_l$ as a function of temperature $T$.
Data are connected to guide the eyes.
If not presented, the error bars are within the size of the data point.
\vspace{0.3 cm}

Fig.~2b. \\
Modified Lindeman ratio $\delta_l$ as a function of quantum parameter $q$
at $T=0.6$.
\vspace{0.3 cm}

Fig.~3a. \\
The dependence of the helicity modulus $\gamma$
(and $\gamma_{cls}$) {\it vs.} quantum parameter
$q$ at temperature $T = 0.5$. $N \times N = 20 \times 20$.
The dependence $\frac{2T}{\pi} = 1 / \pi$ (see in text)
is shown with the help of a dashed line.
\vspace{0.3 cm}

Fig.~3b. \\
The dependence of the helicity modulus $\gamma$ (and $\gamma_{cls}$)
{\it vs.} temperature $T$ at different quantum parameters $q$:
Insert: $q=0.5$ ($N \times N = 10 \times 10$);
\vspace{0.3 cm}

Fig.~4a. \\
The vorticity modulus $v_{DAP-P}$ (\ref{Vorticity}) as a function of
temperature $T$ at different values of the quantum parameter.
$N \times N = 12 \times 12$).\\
Open squares are
the empirical coefficient $\xi = \xi(T)$ in the expression for the
space correlation function of phases $g(r) \sim \exp(-r / \xi)$;
$q=0.7$, $N \times N = 20 \times 20$.
\vspace{0.3 cm}

Fig.~4b. \\
The vorticity modulus $v_{DAP-P}$ {\it vs.} quantum
parameter $q$ at $T=0.5$, $N \times N = 12 \times 12$.
Open squares are
the empirical coefficient $\xi = \xi(q)$ (see in text) in the expression
for the space correlation function of phases (\ref{Corr_function}) at
$T=0.5$, $N \times N = 20 \times 20$.
\vspace{0.3 cm}

Fig.~5a. \\
The vorticity modulus $v_H$ (\ref{v}) as a function of
temperature $T$ at $q=1.5$.
\vspace{0.3 cm}

Fig.~5b. \\
The vorticity modulus $v_H$ (\ref{v}) as a function of $q$ at $T=0.6$.
\vspace{0.3 cm}


\begin{references}
\bibitem[*)]{e-mail} e-mail: lozovik@isan.troitsk.ru

\bibitem{Abeles}  B. Abeles, Phys. Rev. B, {\bf 15}, 2828 (1977).

\bibitem{Efetov}  K.V. Efetov, JETP, {\bf 78}, 2017 (1979).

\bibitem{Doniach} S. Doniach, Phys. Rev. B, {\bf 24}, 5063 (1981).

\bibitem{Akopov}    Yu.E. Lozovik, S.G. Akopov, {\it J.Phys.C} {\bf 14},
                    L31, (1981);
                    S.G. Akopov, Yu.E. Lozovik, {\it J.Phys.C} {\bf 15}, 4403
                    (1982).

\bibitem{Imry}    D. Imry, Phys. Rev. B, {\bf 24}, 6353 (1985).

\bibitem{Zaikin}  G. Sch\"on, A.D. Zaikin, Phys. Rep. {\bf 198}, 5\&6,
                    237 (1990).

\bibitem{Choi}      J.B. Kim, M.Y. Choi, {\it Phys.Rev.B} {\bf 52}, 3624, (1995).

\bibitem{JosePhRev} L. Jacobs, J.V. Jose and M.A. Novotny,
                    {\it Phys.Rev.Lett.} {\bf 26}, 2177, (1984);
                    L. Jacobs, J.V. Jose, M.A. Novotny and A.M. Goldman,
                    {\it Phys.Rev.B} {\bf 38}, 4562, (1988);
                    C. Rojas, J.V. Jose,
                    Phys. Rev. B, {\bf 54} 12361 (1996).

\bibitem{Kultanov}  N.K. Kultanov, Yu.E.Lozovik, {\it Sol.St.Comm.} {\bf 88},
                    645, (1993); N.K. Kultanov, Yu.E.Lozovik,
                    {\it Phys.Lett.A} {\bf 198}, 165, (1995).

\bibitem{Hikami}    S. Hikami and T. Tsuneto, {\it Prog.Theor.Phys.}
                    {\bf 63}, 387, (1980).

\bibitem{Bruder} C. Bruder, R. Fazio et al.
                 {\it Phys. Scr.} {\bf T 42}, 159, (1992).

\bibitem{BelousJPC} A.I. Belousov, S.A. Verzakov and Yu.E. Lozovik
                    {\it J. Phys.: Cond. Matt.} {\bf 10}, 1079, (1998);
                    A.I. Belousov, Yu.E. Lozovik
                    {\it JETP. Lett.} {\bf 66}, 25, (1997);

\bibitem{Ceperley}  D.M. Ceperley, {\it Rev.Mod.Phys.} {\bf 67},
                    279, (1995).

\bibitem{Sayer}     W. Janke, T.Sayer, {\it Chem.Phys.Lett.} {\bf 201},
                    499, (1993).

\bibitem{Bedanov}   Yu.E. Lozovik, V.M. Farztdinov, {\it Sol. St. Commun.},
                    {\bf 54}, 725, (1985);
                    V.M.Bedanov, G.V.Gadiyak, Yu.E.Lozovik,
                    {\it Phys. Lett.} {\bf 109A}, 289, (1985).

\bibitem{Ohta}      M.E. Fisher,  M.N. Barber and  D. Jasnow,
                    {\it Phys.Rev.A}  {\bf 8},  1111,  (1973);
                    T. Ohta,  D. Jasnow,
                    {\it Phys.Rev.B} {\bf 20}, 139, (1979).

\bibitem{Minhagen}  P. Minnhagen, {\it Rev.Mod.Phys.} {\bf 59}, 1001, (1987);
                    T. Ohta,  D. Jasnow,
                    {\it Phys.Rev.B} {\bf 20}, 139, (1979).

\bibitem{Papricash} S. Teitel, C. Jayaprakash. {\it Phys.Rev.B}
                    {\bf 27}, 598, (1983).

\bibitem{Shenoy}    G. Williams, {\it Phys.Rev.Lett.} {\bf 59}, 1926, (1987);
                    S.R. Shenoy, {\it Phys.Rev.B}
                    {\bf 40}, 5056, (1989) and references therein.

\bibitem{Kicuchi}   W. Kawamura, H. Kikuchi, {\it Phys.Rev.B} {\bf 47},
                    1134, (1993).

\bibitem{Hogerson}  G.J. Hogerson, W.P. Reinhardt, {\it J.Chem.Phys.} {\bf 102},
                    4151, (1995).

\bibitem{Southern} B.W. Southern, H-J. Xu, Phys. Rev. B, {\bf 52}, R3836
                   (1995).

\bibitem{Thouless}  J.M. Kosterlitz, D.J. Thouless, {\it J.Phys.C}
                    {\bf 6}, 1181, (1973).

\bibitem{Verzak}    Yu. E. Lozovik, S.A. Verzakov,
                    {\it Phys. of the Sol. State (Fizika Tverdogo Tela)}
                    {\bf 39}, 818, (1996).
\end{references}
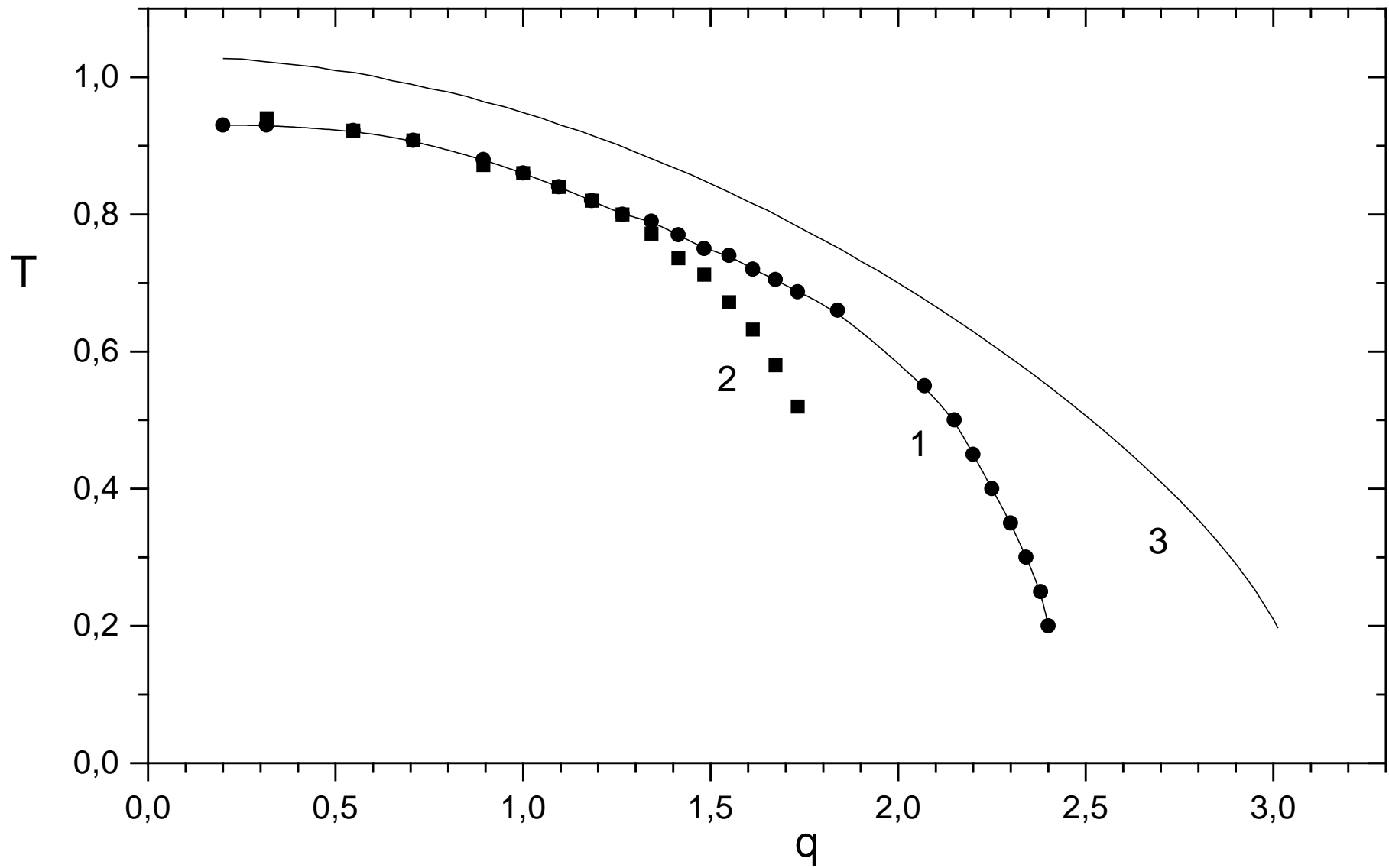
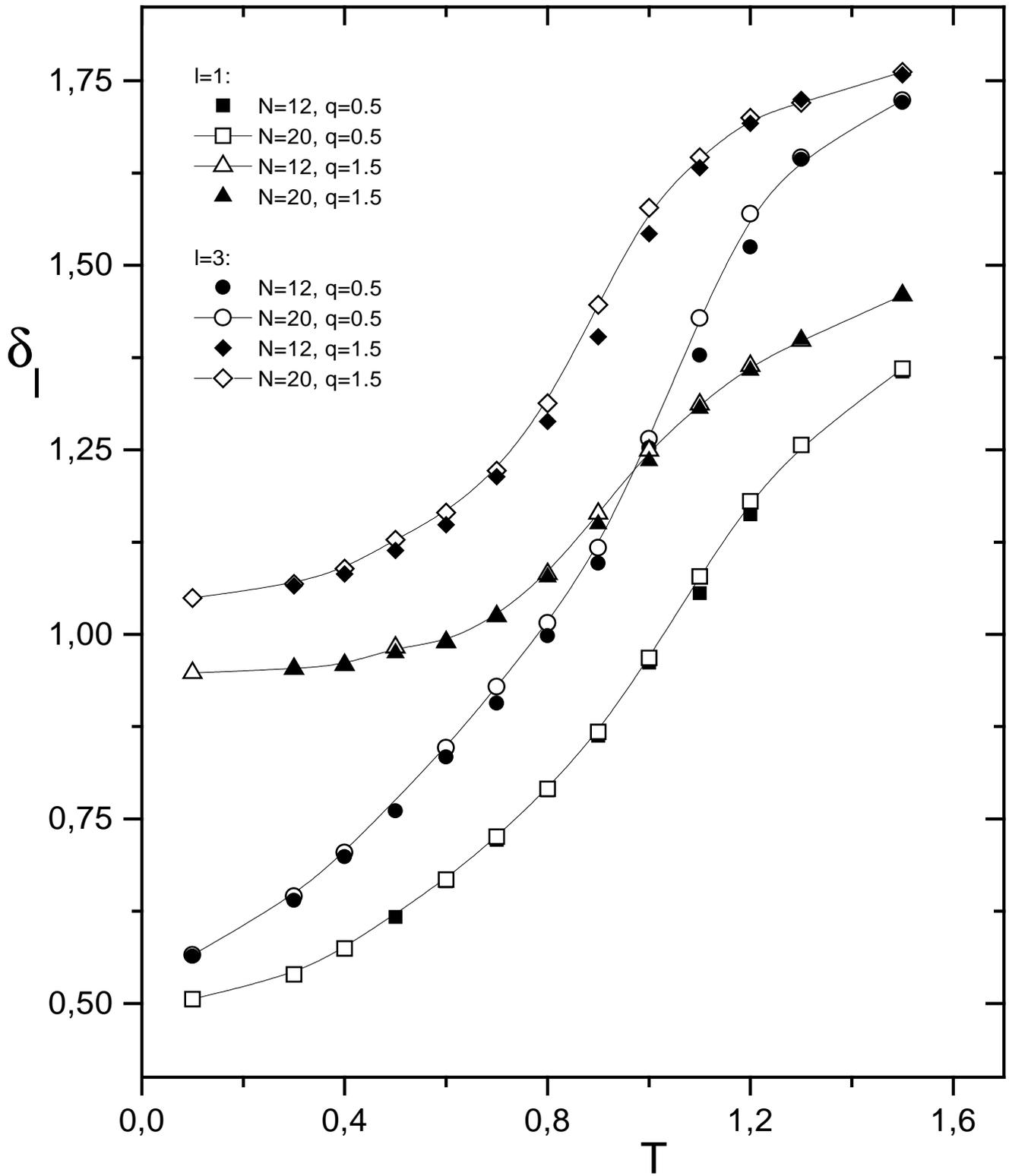
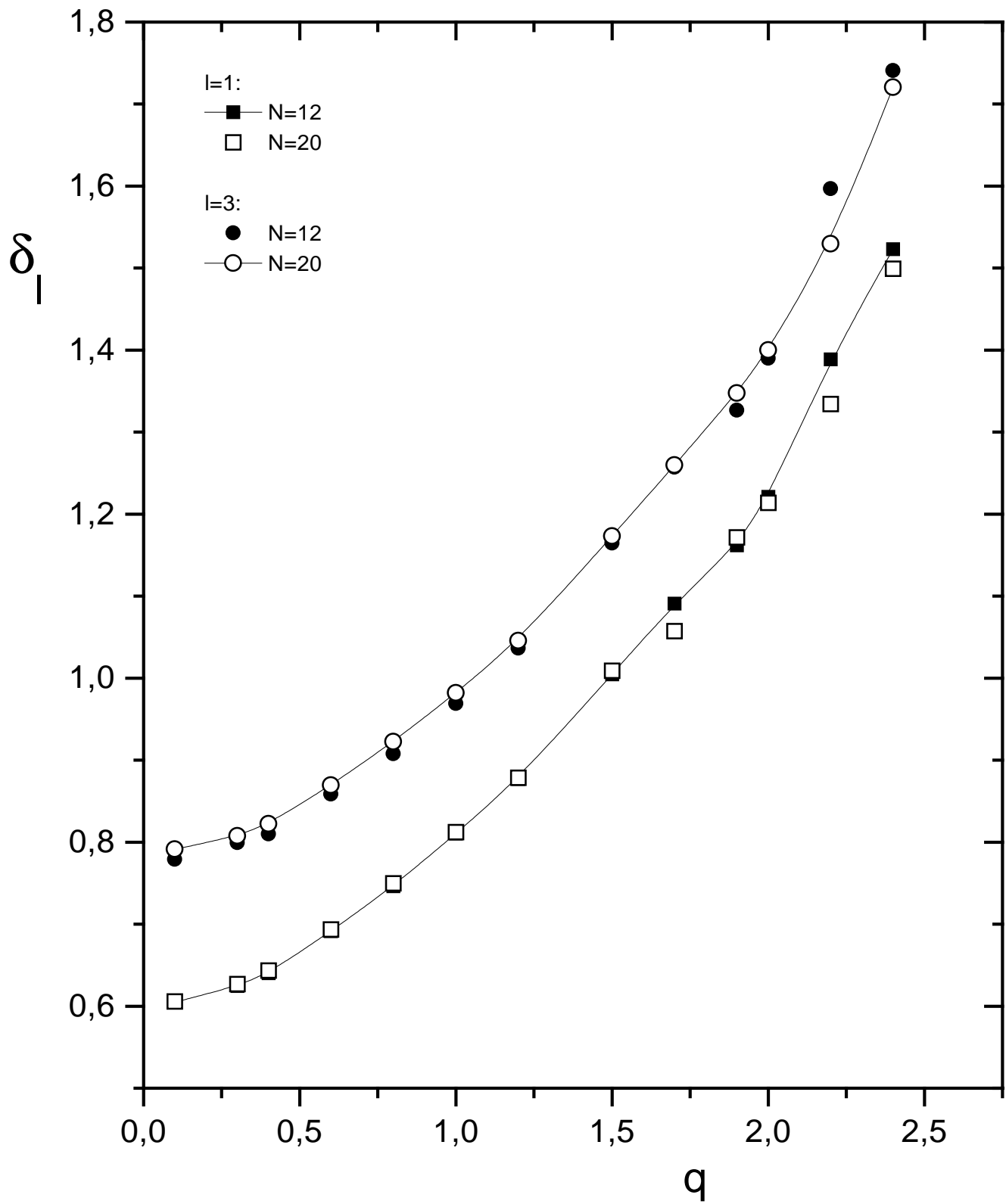
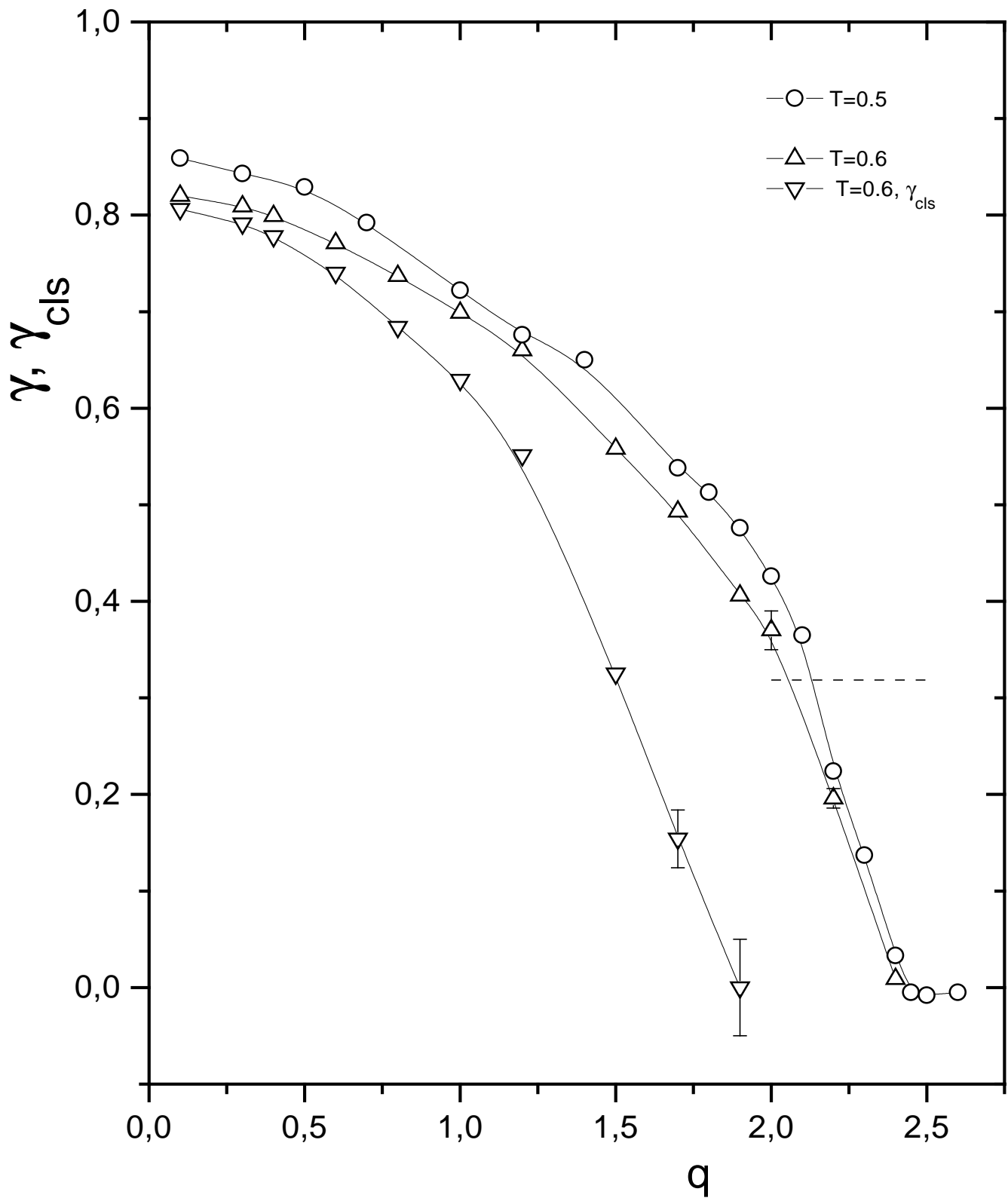
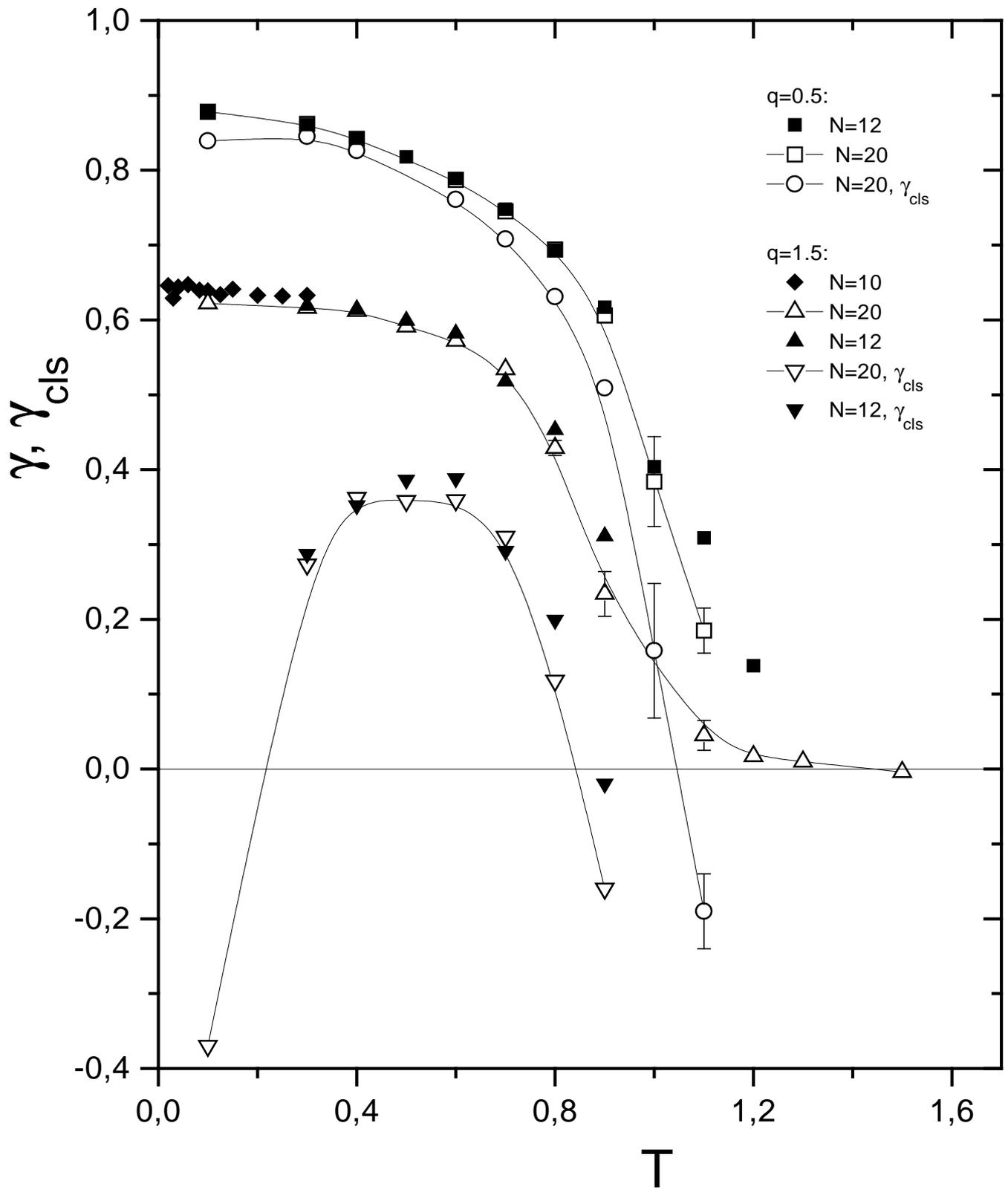
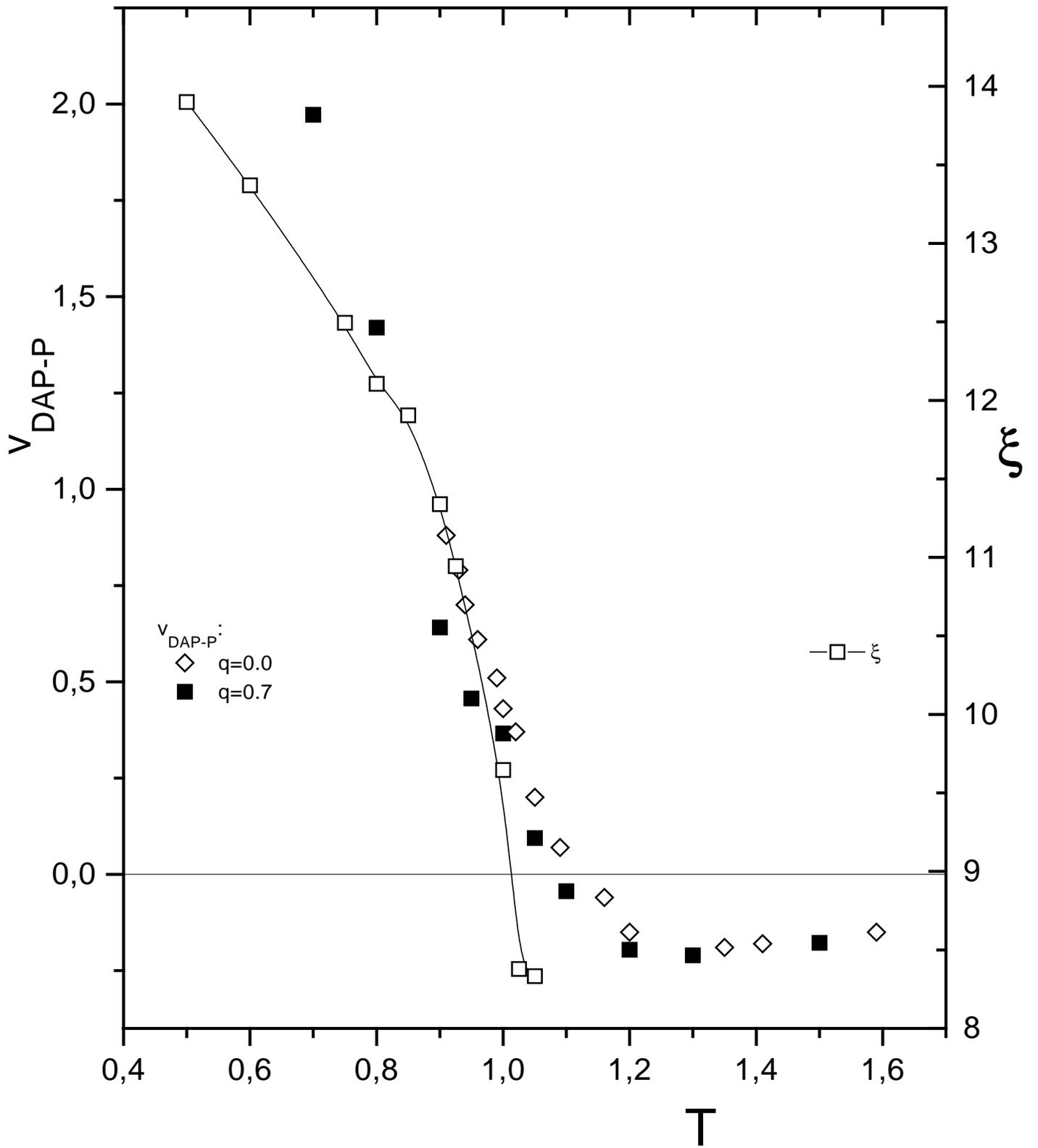
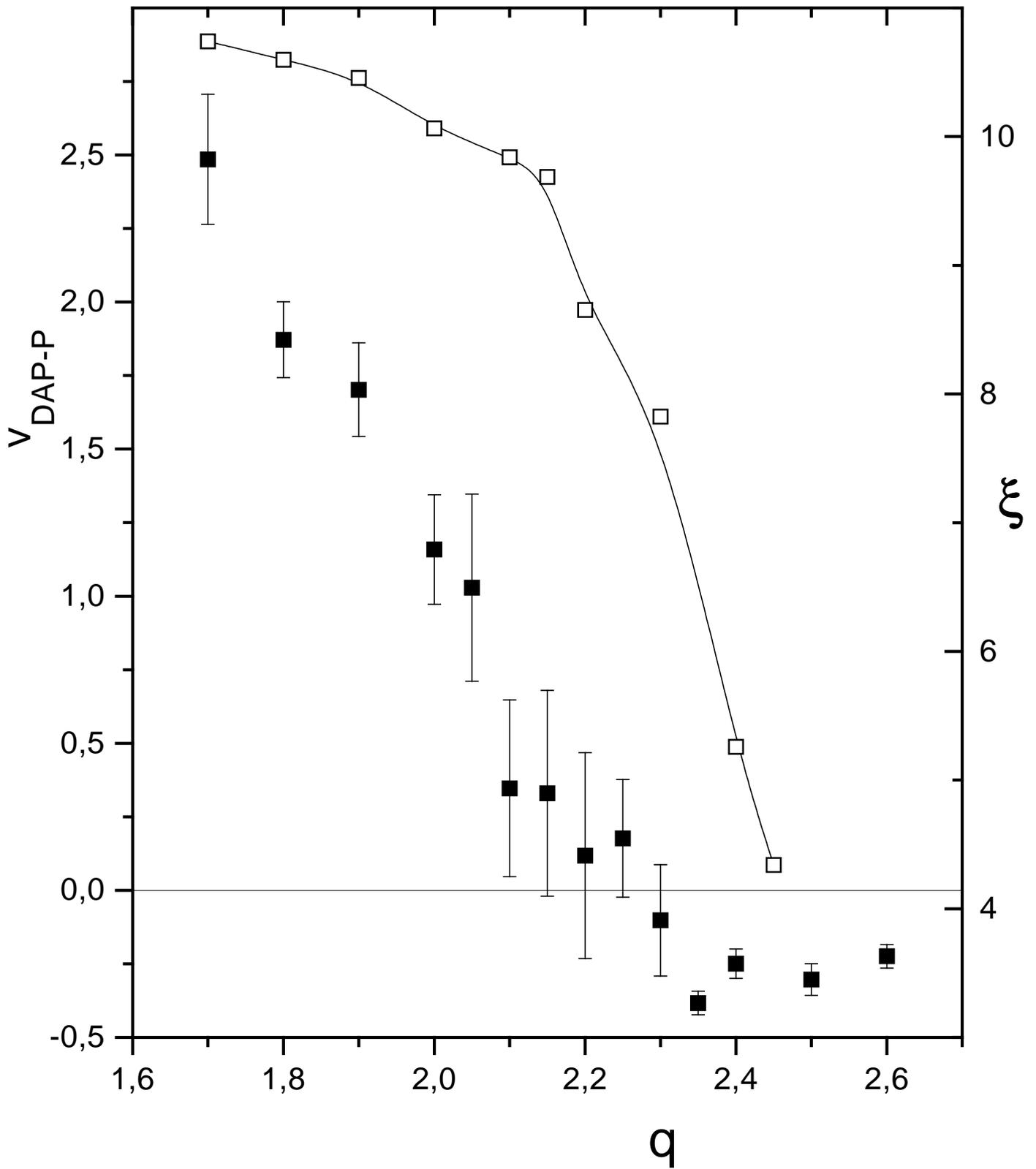
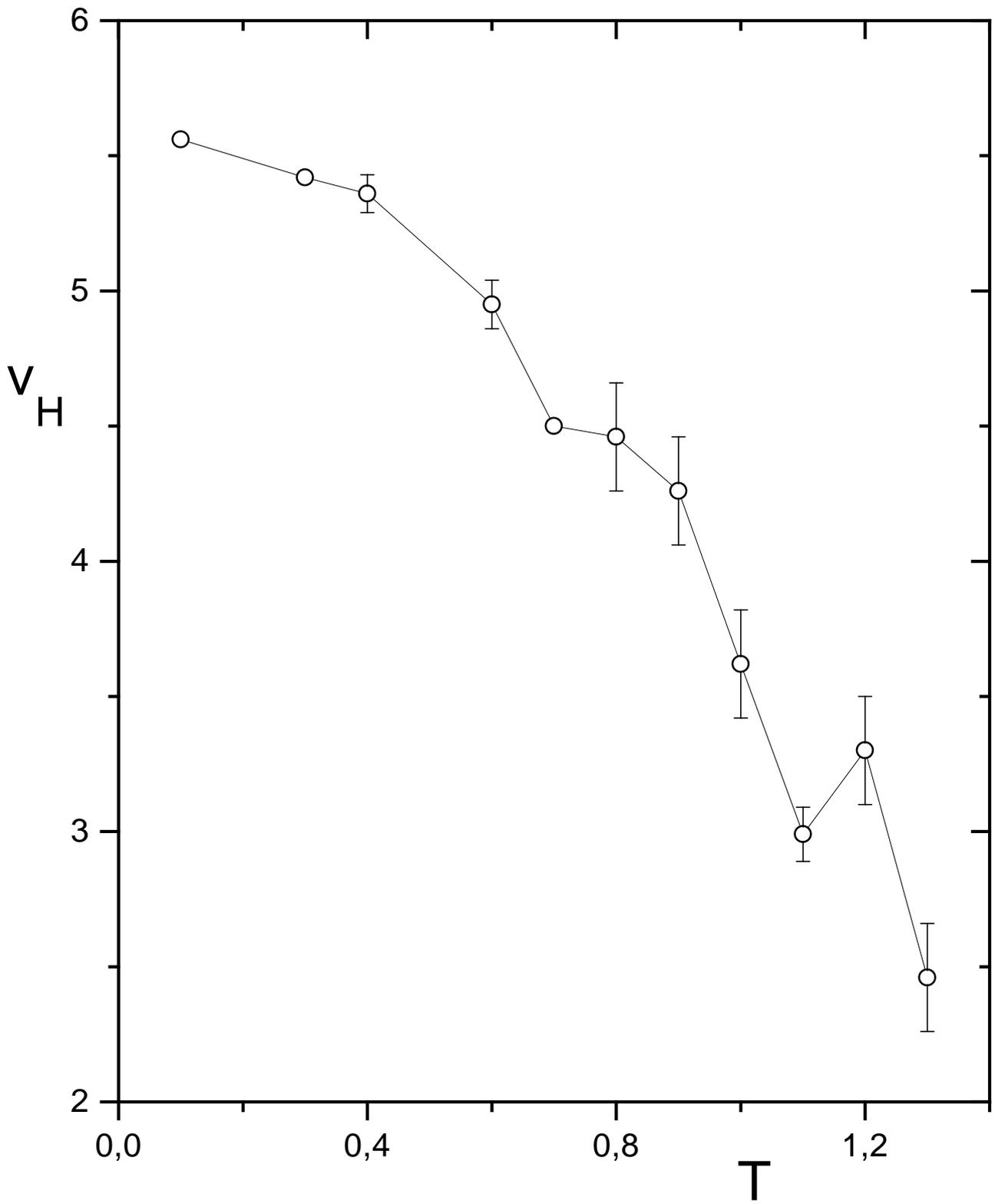
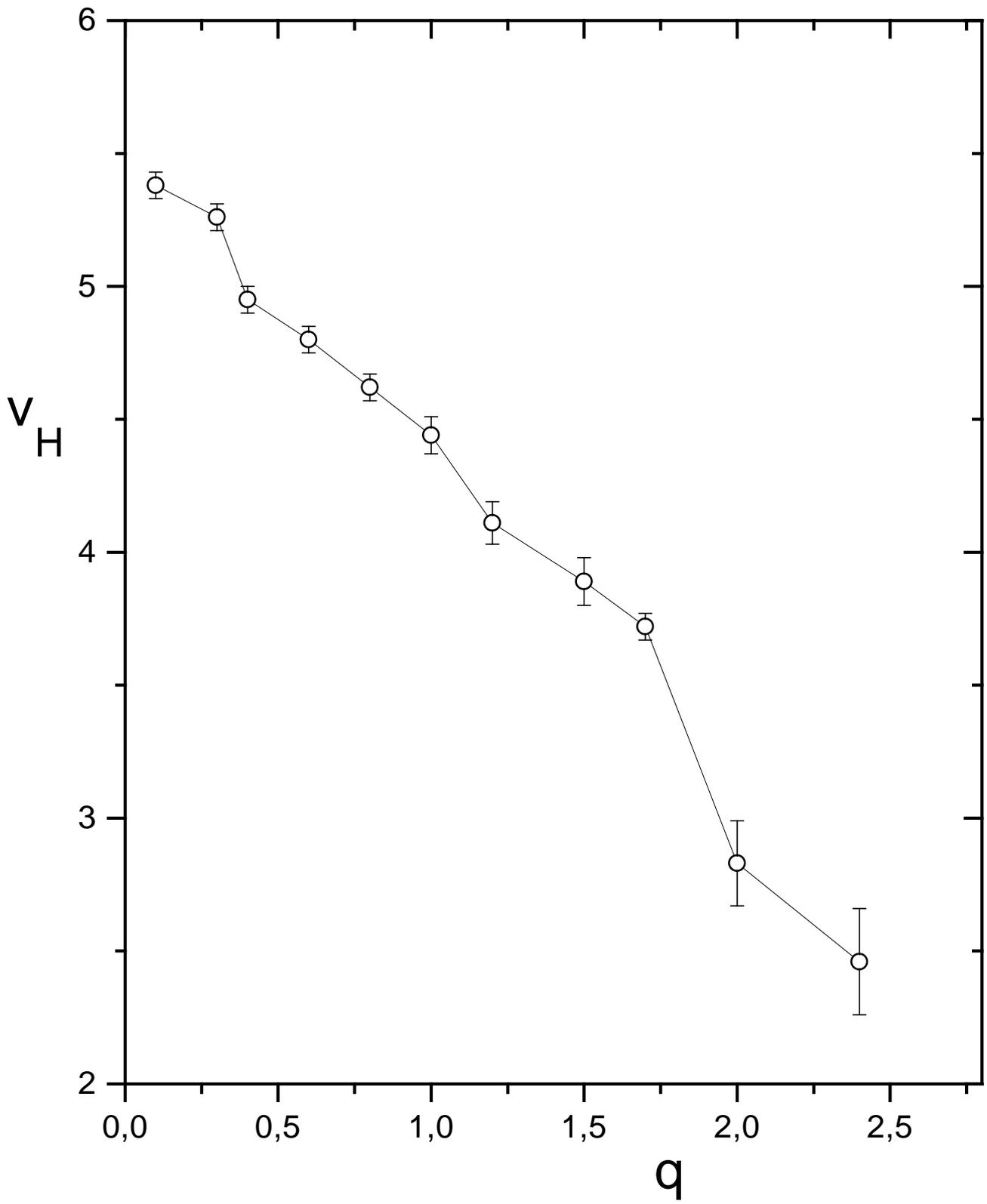
\end{document}